\documentclass[aps,prl,reprint,floatfix,superscriptaddress]{revtex4-2}
\usepackage{blindtext}
\usepackage{graphicx}
\usepackage{braket}
\usepackage{color}
\usepackage{siunitx}
\usepackage{amsmath}
\usepackage{tabularx}
\usepackage{xspace}
\usepackage{amssymb}
\usepackage{statmath}
\usepackage{bm}
\usepackage{ragged2e}
\usepackage{subcaption}

\newcommand{\HCI}[2]{#1$^{#2+}$\xspace}
\newcommand{\Ni}[0]{\HCI{Ni}{12}}
\newcommand{\Be}[0]{\HCI{Be}{}}

\newcommand{\threePtwo}[0]{$^3\!P_2$\xspace}
\newcommand{\threePone}[0]{$^3\!P_1$\xspace}
\newcommand{\threePzero}[0]{$^3\!P_0$\xspace}

\begin{document}
	\title{Finding the ultra-narrow \threePtwo $\rightarrow$ \threePzero electric quadrupole transition in \Ni ion for an optical clock}
    \author{Charles Cheung}
    \author{Sergey G. Porsev}
    \author{Dmytro Filin}
    \author{Marianna S. Safronova}
    \affiliation{Department of Physics and Astronomy, University of Delaware, Newark, Delaware 19716, USA}

	\author{Malte Wehrheim}
	\email{malte.wehrheim@quantummetrology.de}
	\author{Lukas J. Spieß}
	\author{Shuying Chen}
	\author{Alexander Wilzewski}
	\affiliation{Physikalisch-Technische Bundesanstalt, Bundesallee 100, 38116 Braunschweig, Germany}
	
	\author{José R. Crespo López-Urrutia}
	\affiliation{Max-Planck-Institut für Kernphysik, Saupfercheckweg 1, 69117 Heidelberg, Germany}
	
	\author{Piet O. Schmidt}
	\email{piet.schmidt@quantummetrology.de}
	\affiliation{Physikalisch-Technische Bundesanstalt, Bundesallee 100, 38116 Braunschweig, Germany}
	\affiliation{Institut für Quantenoptik, Leibniz Universität Hannover, Welfengarten 1, 30167 Hannover, Germany}

	\begin{abstract}
The Ni$^{12+}$ ion features an electronic transition with a natural width of only 8 mHz, allowing for a highly stable optical clock. We predict that the energy of this strongly forbidden $3s^2 3p^4\, ^3\!P_2 \rightarrow 3s^2 3p^4  \, ^3\!P_0$ electric quadrupole transition is 20081(10) cm$^{-1}$. For this, we use both a hybrid approach combining configuration interaction (CI) with coupled-cluster (CC) method  and a pure CI calculation for the complete 16-electron system, ensuring convergence. The resulting very small theoretical uncertainty of only 0.05\% allowed us to find the transition experimentally in a few hours, yielding an energy of 20078.984(10) cm$^{-1}$. This level of agreement for a 16-electron system is unprecedented and qualifies our method for future calculations of many other complex atomic systems. While paving the way for a high-precision optical clock based on Ni$^{12+}$, our theory and code development will also enable better predictions for other highly charged ions and other complex atomic systems. 
    \end{abstract}
	
	\maketitle
	
	\textit{Introduction.}
Highly charged ions (HCI) are very interesting for searches for physics beyond the Standard Model of elementary particles and interactions~\cite{Kozlov2018RMP}. 
Several of their predicted optical transitions have some of the highest sensitivities to variations of the fine-structure constant and accordingly enhance the corresponding dark matter searches and tests of local position invariance~\cite{Safronova2018RMP,Kozlov2018RMP,2018LI}. Such optical transitions are laser accessible and can provide high-precision frequency standards in atomic clocks~\cite{Kozlov2018RMP,king_optical_2022}. 

Developing highest-accuracy clocks based on HCI is motivated by their extreme atomic properties: a much lower sensitivity to external electromagnetic perturbations than singly charged ions or neutral atoms, and a strong suppression of systematic frequency shifts, such as ac Stark, electric quadrupole, and higher-order magnetic field shifts, and from blackbody radiation.

This new class of clocks was demonstrated in 2022 with the reported optical magnetic-dipole transition in Ar$^{13+}$ with an evaluated systematic frequency uncertainty of $2.2 \times 10^{-17}$, comparable to current operating optical clocks~\cite{king_optical_2022, ludlow_optical_2015}. Increasing this accuracy is unpractical with Ar$^{13+}$ due to the relatively short lifetime of the excited state, limiting interrogation times and, accordingly, the statistical uncertainty of optical frequency comparisons due to quantum projection noise. Thus, we have chosen the strongly forbidden $3s^2 3p^4 \, ^3\!P_2 \rightarrow 3s^2 3p^4   \, ^3\!P_0$  electric quadrupole clock transition in Ni$^{12+}$ \cite{2018Ni} as the most suitable candidate for the next HCI clock, with a convenient wavelength, narrow transition, relatively simple electronic structure, and expected small systematic effects. 

The main obstacle here was the large uncertainty of the clock state excitation energy of 0.5\% (\SI{100}{\per\centi\meter} or \SI{3}{\tera\hertz}), which would have required long scan times to find the extremely weak transition, which has a linewidth of only \SI{8}{\milli\hertz}. 
In contrast, the wavelength of the $^3\!P_1 \rightarrow \, ^3\!P_0$ $M1$ transition could be measured in emission due to its higher transition rate, resulting in a more accurate value \cite{2024Ni} 19541.758(18) cm$^{-1}$. In addition, two more energy levels of the same configuration are reasonably well known and serve as computational benchmarks. In this Letter, we show how the problem is solved with a uniquely precise computation of Ni$^{12+}$ low-lying energies with an uncertainty below \SI{10}{\per\centi\meter}, unprecedented for such a complicated atomic system. Applying recently developed fast laser frequency scanning techniques \cite{2024search}, our prediction allowed us to locate the sought-after transition within just one day at \SI{20078.984\pm0.010}{\per\centi\meter}, only \SI{2}{\per\centi\meter} away from it. Our work has thus paved the way for the development of the Ni$^{12+}$ clock and demonstrates the reliability of future calculations needed for the search of other clock transitions in HCIs.
\begin{table*}
\caption{\label{bspline_r7} \justifying Contributions to the excitation energies of the low-lying states (in cm$^{-1}$) calculated using the CI+all-order method. 
See the main text for detailed explanations of all contributions.  Experimental values are displayed in column ``Expt.~\cite{NIST_ASD,2024Ni}''. The last two columns display the difference between the theoretical and experimental values.
} 
\begin{ruledtabular}
\begin{tabular}{lcccccccccccccc}
\multicolumn{1}{c}{Configuration}&
\multicolumn{1}{c}{$17spdf\!g$}& 
\multicolumn{1}{c}{18-22}& 
\multicolumn{1}{c}{23-29}& 
\multicolumn{1}{c}{6-28}& 
\multicolumn{1}{c}{7-28}& 
\multicolumn{1}{c}{$l>6$}&
\multicolumn{1}{c}{Extra}&
\multicolumn{1}{c}{NL+Tr}&
\multicolumn{1}{c}{TEI}&
\multicolumn{1}{c}{QED}&
\multicolumn{1}{c}{Total}&
\multicolumn{1}{c}{Expt.~\cite{NIST_ASD,2024Ni}}&
\multicolumn{1}{c}{Diff.}&
\multicolumn{1}{c}{Diff. (\%)}\\
\multicolumn{2}{c}{}&
\multicolumn{1}{c}{$spdf\!g$}& 
\multicolumn{1}{c}{$spdf\!g$}&
\multicolumn{1}{c}{$h$}&
\multicolumn{1}{c}{$i$}&
\multicolumn{1}{c}{}&
\multicolumn{1}{c}{conf.}&
\multicolumn{7}{c}{}\\
\hline \\[-0.7pc]
$3s^2\,3p^4\,^3\!P_1$ &  19459 &   9 &  0&   8&4&  4   &  0 & -3    & 16  & 50&  19547 & 19542 &  6 & 0.03\% \\
$3s^2\,3p^4\,^3\!P_0$ &  20114 &  -3 & -1&  -8&-4& -4  &  -5& -3    & -34 & 33&  20086 & 20060(100) &  & \\
$3s^2\,3p^4\,^1\!D_2$ &  47441 & -49 &  4& -98&-46&-40 & -19& -2    &-163 & 43&  47063 & 47033 & 30 & 0.06\% \\
$3s^2\,3p^4\,^1\!S_0$ &  98686 & -89 &  9&-134&-66&-64 & -56&  -8   &-416 & 50& 97894 & 97836 & 58 & 0.06\% \\
  \end{tabular}
\end{ruledtabular}
\end{table*}

\textit{Theory: CI+all-order method.} 
Ni$^{12+}$ has 16 electrons in its ground state $1s^2 2s^2 2p^6 3s^2 3p^4$ configuration. 
 We start with a hybrid approach that combines the configuration interaction (CI) method with the linearized coupled-cluster single-double (all-order) method \cite{SafKozJoh09}.
 In our ``CI+all-order'' method, CI is used to treat the six outer $3s^2 3p^4$ electrons, which we consider to be valence electrons. To account for the correlations between the ten core $1s^2 2s^2 2p^6$ and valence electrons, we applied the coupled-cluster method as described in Ref.~\cite{SafKozJoh09}. 
 
\begin{figure}
	\centering
	\includegraphics[width=\columnwidth]{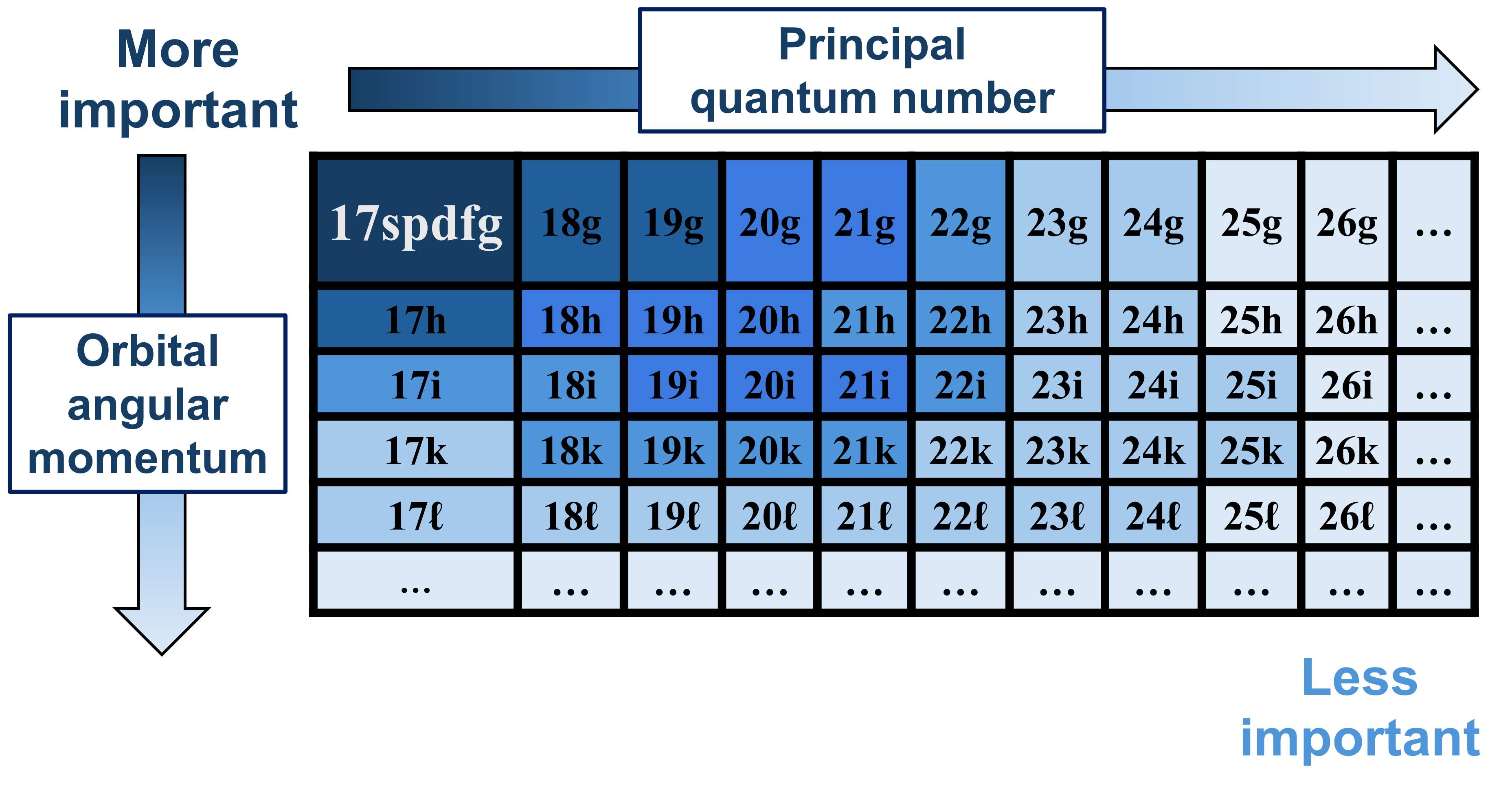}
	\caption{\label{fig1} \justifying Illustration of the basis set upscale. $17spdf\!g$ includes all orbitals up to $n=17$ for $spdf\!g$ partial waves. Darker blocks indicate orbitals with larger contributions to the energies and wave functions.}
	\end{figure}
The CI wave function is obtained as a linear combination of all distinct states of a given angular momentum $J$ and parity
$
    \Psi_J=\sum_ic_i\Phi_i.
$
In the pure CI approach, the wave functions and energies of the low-lying states are obtained by solving the many-electron Schr\"odinger equation $H \Phi_n = E_n \Phi_n$~\cite{KozPorFla96}.
The CI+all-order approach incorporates core excitations in the CI method by defining an effective Hamiltonian 
$
    H_\mathrm{eff}(E)=H+\Sigma(E),
$
where $H$ is the Hamiltonian in the frozen-core approximation. The energy-dependent operator $\Sigma(E)$ is constructed using the coupled-cluster method to account for the virtual excitations of the ten core electrons \cite{SafKozJoh09}. 

The one-electron orbitals of the core shells were obtained by solving the Dirac-Hartree-Fock equations in the central field approximation using a $V^{N-6}$ basis, where $N$ is the number of electrons. 
The basis set was constructed on a radial grid using $B$ splines, with 40 splines of order 7, constrained to a spherical cavity of radius $R=7$ a.u. This included a total of seven partial waves ($l_\mathrm{max}=6)$ and orbitals with a principal quantum number $n$ of up to 35. The Breit and Coulomb interactions were included on the same footing for constructing the basis set, which is large enough for the coupled-cluster computations to accurately generate the effective Hamiltonian.


We apply several approximations of increasing accuracy to construct the effective Hamiltonian: second-order many-body perturbation theory (MBPT), linearized coupled-cluster method with single and double excitations (LCCSD), and coupled-cluster method with single, double, and triple excitations (CCSDT). 
We find very small differences, from -1~cm$^{-1}$ for the $^3\!P_1$ state to -10~cm$^{-1}$ for the $^1\!S_0$ state between the CI+LCCSD and the CI+MBPT results, showing that all higher-order 
core-valence correlations are very small. The CCSDT method included non-linear single and double terms and linear triple terms, which are small and are listed under the column ``NL+Tr'' in Table~\ref{bspline_r7}. 

The convergence of the 6-electron CI calculation to an accuracy of a few cm$^{-1}$ is a major computational challenge and had not been demonstrated prior to this work to our knowledge. 
We start with all possible single and double (SD) excitations from the $3s^2 3p^4$, $3s^2 3p^3 4p$, $3s 3p^4 4s$, $3s 3p^4 3d$, $3s^2 3p^2 3d^2$, and $3p^6$ reference configurations to 
$17spdf\!g$, which means that
all orbitals with the principal quantum number up to $n=17$ and the $spdf\!g$ partial waves were included. The resulting energies are listed in Table~\ref{bspline_r7} in the column ``$17spdf\!g$.'' We expand the basis set and increase the number of starting configurations until a numerical convergence is reached in all parameters of the CI calculation.
The procedure for extending the basis set is illustrated in Fig.~\ref{fig1}. We start from the six basic configurations mentioned above, but perform SD excitations to growing sets of orbitals. First, we only increase the principal quantum number to $n=29$ while keeping the maximum orbital angular momentum $l=4$ (i.e., $spdf\!g$ partial waves). The resulting contributions listed in Table~\ref{bspline_r7} show that convergence has been reached for the $spdf\!g$ partial waves, given a negligible contribution of states with $n>22$ for the $^3\!P_0$ and $^3\!P_1$ levels. 
 \begin{table*}
\caption{\label{r5} \justifying Contributions to the excitation energies of the low-lying states (in cm$^{-1}$) calculated using the 16-electron CI method. Results in the three columns labeled ``6 electrons'' are obtained, allowing excitations from the last 6 electrons.  In all other calculations labeled ``All electrons'', the excitations are allowed from all 16 electrons. See text for detailed explanations of all contributions. Experimental values are displayed in column ``Expt.~\cite{NIST_ASD,2024Ni}''. 
} 
\begin{ruledtabular}
\begin{tabular}{lcccccccccccc}
\multicolumn{1}{c}{}&
\multicolumn{3}{c}{6 electrons}&
\multicolumn{4}{c}{All electrons}&
\multicolumn{5}{c}{}
\\
\multicolumn{1}{c}{Configuration}&
\multicolumn{1}{c}{$17$}& 
\multicolumn{1}{c}{$17$}& 
\multicolumn{1}{c}{18-20}& 
\multicolumn{1}{c}{$11$}&
\multicolumn{1}{c}{12-18}&
\multicolumn{1}{c}{Extra}&
\multicolumn{1}{c}{Estimate}&
\multicolumn{1}{c}{QED}&
\multicolumn{1}{c}{Total}&
\multicolumn{1}{c}{Expt.~\cite{NIST_ASD,2024Ni}}&
\multicolumn{1}{c}{Diff}&
\multicolumn{1}{c}{Diff (\%)}\\
\multicolumn{1}{c}{}&
\multicolumn{1}{c}{$spdf\!g$}&
\multicolumn{1}{c}{$l>4$}& 
\multicolumn{1}{c}{$spdf\!ghikl$}&
\multicolumn{1}{c}{$spdf\!g$}&
\multicolumn{1}{c}{$spdf\!g$}&
\multicolumn{1}{c}{conf.} &
\multicolumn{1}{c}{full CI} &
\multicolumn{4}{c}{}
\\
\hline \\[-0.7pc]
$3s^2\,3p^4\,^3\!P_1$ & 19395 &  16 &0&   83 &    6 &   1 &   0 & 49 &  19550 & 19542 &   8 &  0.04\% \\
$3s^2\,3p^4\,^3\!P_0$ & 20052 & -15 &0&   14 &   -5 &   5 &  -3 & 33 &  20081 & 20060(100) &  &   \\
$3s^2\,3p^4\,^1\!D_2$ & 47348 & -179 &-1& -105 &   -2 & -45 &  -5 & 42 &  47051 & 47033 &  18 &  0.04\% \\
$3s^2\,3p^4\,^1\!S_0$ & 98397 & -260 &-4& -318 & -110 &  38 & -24 & 52 &  97771 & 97836 & -66 & -0.07\% \\
  \end{tabular}
\end{ruledtabular}
\end{table*}

Next, in addition to all excitations previously included~\cite{comment}, we allow SD excitations to the orbitals of $h$, $i$, $k$, and $l$ partial waves. We list the contributions 
of the $h$ and $i$ partial waves separately to illustrate their significance. Usually, CI computations include only the first few orbitals, for example, 6-12$h$, for high partial waves. We unexpectedly find that this practice drastically underestimates their contribution. In fact, more than half of the contribution for the $h$ partial wave comes from the (18-28)$h$ orbitals in this basis set. This trend worsens for the $i$ orbitals, where almost all of the contribution comes from $n>17$. In the pure CI computation described below, we can partly remedy this issue by using a smaller, more compact basis built using a recurrent procedure with a better $V^{N}$ potential. However, these orbitals lie very high in the spectrum, and excitations to orbitals with large $n$ have to be included to ensure convergence.
We extrapolated the contributions of the partial waves with $l>6$ based on the convergence of energies for the $h$, $i$, $k$ and $l$ partial waves. In Table~\ref{bspline_r7}, we list the total contribution to the energies of higher partial waves in the column ``$l>6$.''

Further significant contributions due to excitations from additional reference configurations are listed in column ``Extra conf.'' in Table~\ref{bspline_r7}. 
To select a new reference configuration, we order all configurations from the $17spdf\!g$ run by their weights, and then extend the set of the reference configurations by adding those with the largest weight first. We first allow single excitations from 1200 reference configurations, then SD excitations from 41 more, and keep adding configurations until we reach convergence with good numerical accuracy.
We note that CI convergence required a tremendous increase of the number of configurations: single-double excitation to the $17spdf\!g$ orbitals produces 256,000 configurations; expanding the basis set to $28spdf\!gh$ yields another 933,000 configurations, while including $i$ orbitals adds another 418,000 configurations. A total of 3.3 million extra configurations were included to obtain the result in column ``Extra conf.'' These calculations became possible through improvements of the highly parallel pCI package \cite{2021sym, 2025pCI}, and separation of the total computations of more than 5 million configurations into a large number of manageable computations using Python interfaces developed to automate this process.  
 
We calculated QED corrections and three-electron interaction (TEI) corrections following Refs.~\cite{QED} and~\cite{KozSafPor16}, respectively, and listed them in columns ``QED'' and ``TEI''.  The adequate (10\% or better) accuracy of QED calculations has been established in Ref.~\cite{2024QED}. TEI corrections are due to our separation of the computations into core and valence sectors and are not accounted for by the CI+all-order method. 
In column ``Total'', we present the final energies, calculated by adding all corrections beyond the initial results $17spdf\!g$. In the last two columns of Table~\ref{bspline_r7}, we show the small differences between our final CI+all-order values and the experimental results compiled in the NIST~\cite{NIST_ASD, 2024Ni} database of only 0.03\% for the $^3\!P_1$ state and 0.06\% for the $^1\!D_2$ and $^1\!S_0$ states, respectively. 

    \textit{Theory: pure CI method.}
The largest uncertainty in the previous approach comes from interactions of four and more electrons that are omitted in the TEI-type correction.
Their effect would be intrinsically included in the pure CI approach. Therefore, we now correlate all 16 electrons on the same footing within the CI instead of using the coupled-cluster effective Hamiltonian approach. The challenge now is that the number of configurations needed would require memory resources beyond those commonly available at high-performance clusters. We have investigated different ways of efficiently breaking the CI computation to separate expansions yielding additive energy corrections that enabled us to perform this computation.

First, we form a smaller basis set using the recurrent procedure described in Refs.~\cite{KozPorFla96,KozPorSaf15}, because we no longer need the large basis used for coupled-cluster computations. This basis set is constructed in a more compact radial cavity of 5 a.u., which led to an accelerated convergence with the principal quantum number $n$. 
We start by allowing only SD excitations from the outer six electrons to $20spdf\!ghikl$ orbitals and extrapolating for higher partial waves. These results are listed in the three columns labeled ``6 electrons'' 
in Table~\ref{r5}. We find that CI saturation was achieved already at $n=17$, as the contribution of $n=$18-20 was negligible for all partial waves. 
Then, we allow additional SD excitations from the $1s$, $2s$, and $2p$ shells of the inner ten electrons to the $spdf\!g$ partial waves from the same dominant reference configurations (columns ``All electrons'' of Table~\ref{r5}).
We also verified that additional excitations 
of the inner ten electrons to the
basis states in $l>4$ partial waves gave a negligible contribution.

Next, we analyzed how energy levels change when we add excitations from additional reference configurations with the highest weights, which we select as in the CI+all-order approach described above (see column ``Extra conf.'' in Table~\ref{r5}).

In total, we computed contributions from 9.6 million configurations broken down into 54 separate computations, with our memory of 31~TB limiting the set size for each single run to roughly 350,000 configurations. This work prompted us to develop a neural network-based algorithm that reduced memory needs and computation time by a factor of three, and human involvement even more \cite{2024NN}. The network was trained to select the most important configurations more efficiently in an automated, iterative process. The essentially exact computations above confirmed these neural network results. 

In addition, we estimate the contribution of all remaining configurations in the CI expansion, which is impossible to compute directly. We studied how the difference in weights of a configuration contributing to the excited and ground states affects the respective energy contribution if we consider this configuration as a reference one. We found a linear dependence
(similar for all four levels) with a relative weight of 0.01\% leading to a correction of about \SI{-11}{\per\centi\meter}.

We tested this relation using contributions to the energies of configurations that were already computed. We added weights from 200, 2000, and 9500 additional reference configurations until convergence was reached and estimated the total contribution based on the above relation. We note that 9500 reference configurations would lead to probably billions of configurations in the CI expansion, making it completely untractable with any computational facilities. 
This contribution is listed in column ``Estimate full CI''. To the best of our knowledge, such a quantitative estimate to complete CI has never been made before.  

The QED corrections are listed in the column labeled ``QED'', and the final energies under column ``Total'' are compared with the experimental values \cite{NIST_ASD,2024Ni} displayed in the last two columns. The $^3\!P_0$ energies obtained by two approaches differ by only \SI{5}{\per\centi\meter}. The difference from the experiment for the $^3\!P_1$ state is \SI{6}{\per\centi\meter} and \SI{8}{\per\centi\meter}. Based on this, we conservatively estimate the energy uncertainty for $^3\!P_0$ to be \SI{10}{\per\centi\meter}. We take the CI value \SI{20081\pm10}{\per\centi\meter} to be final since the convergence of the CI was demonstrated and all corrections except QED nearly cancel for the $^3\!P_0$ level.

\textit{Experiment: finding the clock transition.}
Starting with this prediction, we experimentally searched for the $^3\!P_2 \rightarrow\, ^3\!P_0$ clock transition in Ni$^{12+}$. 
For this, we produced HCI in an electron beam ion trap (EBIT) \cite{micke_ebit_2018} and transported them to a cryogenic Paul trap \cite{leopold_cryogenic_2019}, where we co-trapped one \Ni with a single \Be ion for sympathetic cooling and readout through its coupled motion using quantum-logic-inspired methods \cite{schmidt_2005_spectroscopy, micke_coherent_2020}. 

The uncertainty range of our calculation is \SI{300}{\giga\hertz} (\SI{10}{\per\centi\meter}) and can be scanned within one day of continuous operation using the methods introduced in Ref.~\cite{2024search}. In summary, we used our spectroscopy laser at \SI{498}{\nano\meter} to drive excitations to the \threePzero state, which were then detected by the absence of an excitation signal on the \threePtwo $\rightarrow$ \threePone transition. 
For this, we first identified the \threePtwo $\rightarrow$ \threePone transition using an off-resonant optical dipole force (ODF) \cite{2024search}. 
The ODF arises from the two counter-propagating laser beams detuning from the motional frequency of the ion. As the common laser frequency approaches the electronic resonance ($\lesssim $\SI{10}{\mega\hertz}), the ion motion is coherently excited \cite{wolf_2016_non_destructive, 2024search} without populating the electronically excited state \threePone. The ensuing phonon excitation is detected with high efficiency through the \Be logic ion by mapping motional to spin excitation. The \threePtwo $\rightarrow$ \threePone transition frequency is determined to be \SI{585.847263\pm0.000010}{\tera\hertz} using a commercial wavemeter calibrated with an iodine-stabilized laser at \SI{626}{\nano\meter}. This result agrees well with the EBIT emission measurement~\cite{2024Ni} and the transition frequency calculated above. 

\begin{figure}
    \centering
    \includegraphics[width=\columnwidth]{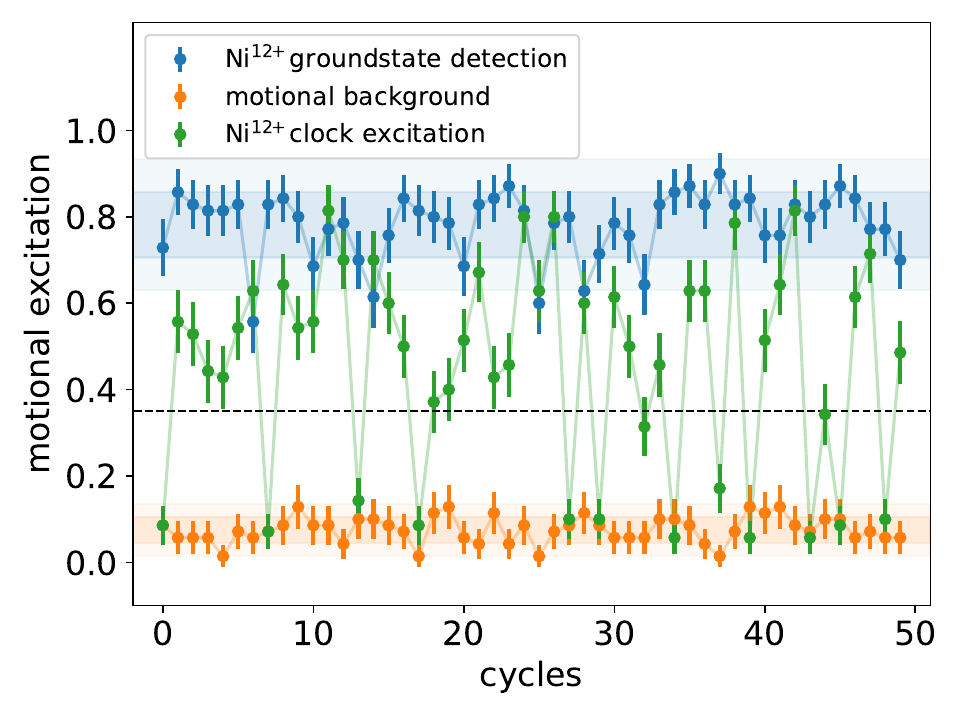}
    
    \caption{\label{fig:clock_excitation} \justifying Measurement of the motional excitation after the \threePtwo $\rightarrow$ \threePzero laser scan. Motional background (orange) and motional excitation (blue) are clearly separated, allowing to threshold the data (black dashed line) for state detection. Scanning across the clock transition resonance frequency (green) leads to repeated quantum jumps appearing in both detection bands (shaded areas) in different scanning cycles. Error bars on the data points display the quantum projection noise from 70 repetitions.}
\end{figure}

In the search for the clock transition, we then used the motion induced by the ODF at a fixed detuning of \SI{5}{\mega\hertz} from the \threePtwo$\rightarrow$\threePone stretched-state resonance as a signal that vanishes upon excitation of the \threePtwo $\rightarrow$ \threePzero clock transition. A frequency-doubled titanium-sapphire laser provided about \SI{100}{\milli\watt} of power at \SI{498}{\nano\meter} for this. 
We estimated an on-resonant Rabi frequency of \SI{20}{\kilo\hertz} and a specified laser linewidth of $<$\SI{100}{\kilo\hertz} averaged over \SI{100}{\micro\second}. 
With these parameters, we estimated an optimal scanning speed of \SI{1}{\giga\hertz\per\second} at 50 cycle repetitions, following Ref.~\cite{2024search} for this experiment.  We expect a 20\% excitation probability to the \threePzero state when the laser frequency hits the transition during a scan. A search cycle consists of three phases. Initially, the \threePtwo, $m_J = 2$ edge state is prepared by optical pumping ($\sim 70$\%) on the \threePtwo $\rightarrow$ \threePone transition, taking approximate \SI{50}{\milli\second}. 
In the second step, the HCI is irradiated by the \SI{498}{\nano\meter} laser, and the laser frequency is linearly swept over a span of \SI{5}{\giga\hertz} in \SI{5}{\second}. Finally, the population of the ground state is read out by interrogating the \threePtwo $\rightarrow$ \threePone transition with the ODF described above. 
The last readout step is repeated 70 times to utilize the experimental dead time of \SI{2}{\second} needed to reset the laser to its sweeping starting frequency. This quantum non-demolition detection scheme \cite{hume_2007_high_fidelity} allowed us to detect the excitation from the ground state with nearly 100$\%$ fidelity. 
In total, we covered in six hours \SI{100}{\giga\hertz} within the uncertainty range until we found the clock transition. Figure~\ref{fig:clock_excitation} shows an example of an off-resonant clock-laser scan leading to high motional excitation by the ODF (blue dots), as well as the motional detection background (orange dots). The shaded color bands cover the region of the mean of the excitation probability and their standard deviations. 
The green points show excitation signals when scanning across the clock transition, and within the background and motional excitation bands provide a clear signature of the clock-transition excitation. 
Off-resonant excitation and cyclic transitions within one scan pump population from the \threePtwo stretched state to other Zeeman sublevels and reduce the motional excitation signal. 
For a more precise measurement, we narrowed the scanning range to \SI{300}{\mega\hertz}, which is our uncertainty from drifts of the not-stabilized titanium-sapphire laser. We determined the \threePtwo $\rightarrow$ \threePzero clock-transition frequency at \SI{601.9528\pm0.0003}{\tera\hertz} (\SI{20078.984\pm0.010}{\per\centi\meter}) with our wavemeter, in excellent agreement with the theory prediction. Future frequency stabilization of the laser to Hz-level linewidth will allow us to perform quantum logic spectroscopy \cite{micke_coherent_2020} and operate an optical clock based on \Ni.  

	\textit{Conclusion.}
We predicted the $^3\!P_2-\,^3\!P_0$ transition energy using two different theoretical approaches, demonstrating full convergence of a 16-electron CI computation including 9.6 million configurations, and developed a method to estimate the residual contributions of the far more numerous configurations that cannot be realistically included in existing computer clusters. This work demonstrates how our advanced theory and code development lead to much smaller calculational uncertainties. Our optimized parallel code package has been made available to the community \cite{2025pCI}. We demonstrated its ability to include millions of configurations by effectively separating different classes of excitations.
Our subsequent development of automated neural network tools further reduces the use of memory and computation time \cite{2024NN}. 

The subsequent experiment demonstrated excitation of this clock transition at \SI{20078.984\pm0.010}{\per\centi\meter}, just \SI{2}{\per\centi\meter} away from the prediction. This remarkable agreement with the \textit{ab initio} calculations of a 16-electron system inspires confidence in future calculations of other complex atomic systems and enables the further development of a Ni$^{12+}$ high-precision atomic clock.
	  
    \begin{acknowledgments}
    We thank Mikhail Kozlov for the helpful discussion of the calculations and Pavlo Bilous for his collaboration on integrating neural network support into CI calculations.
    The project was supported by the Physikalisch-Technische Bundesanstalt, the Max-Planck Society, the Max-Planck–Riken–PTB–Center for Time, Constants and Fundamental Symmetries, and the Deutsche Forschungsgemeinschaft (DFG, German Research Foundation) through SCHM2678/5-2, SU 658/4-2, the collaborative research centers SFB 1225 ISOQUANT and SFB 1227 DQ-\textit{mat}, and under Germany’s Excellence Strategy – EXC-2123 QuantumFrontiers – 390837967. The project 20FUN01 TSCAC has received funding from the EMPIR programme co-financed by the Participating States and from the European Union’s Horizon 2020 research and innovation program. This project has received funding from the European Research Council (ERC) under the European Union’s Horizon 2020 research and innovation program (grant agreement No 101019987). 
    
    The calculations in this work were done through the use of Information Technologies resources at the University of Delaware, specifically the high-performance Caviness and DARWIN computer clusters. 
    The theoretical work has been supported in part by the US NSF Grant  No. PHY-2309254,  US Office of Naval Research Grant No. N00014-20-1-2513, and by the European Research Council (ERC) under the Horizon 2020 Research and Innovation Program of the European Union (Grant Agreement No. 856415).
  \end{acknowledgments}
	\bibliographystyle{apsrev4-2} 
	\bibliography{bibliography.bib} 

\begin{thebibliography}{25}
\bibitem [{com()}]{comment}%
  \BibitemOpen
  \href@noop {} {}\bibinfo {note} {We find the calculations of $k$ and $l$ partial waves to be intractable with such a large basis. We have verified with $h$ and $i$ partial waves that, as expected, the total contribution of a partial wave does not depend on the choice of the basis. We use far more compact basis with the pure CI method to compute the contributions of $k$ and $l$ for partial waves.}\BibitemShut {Stop}%
\end{thebibliography}
	
\end{document}